\documentclass[namedreferences,hyperref,optionalrh]{spr-sola}
\usepackage{graphicx}        
\usepackage{color}           

\chardef\us=`\_

\begin{document}
\begin{frontmatter}

\title{Solar Polar Field Reversals as the Result of the Global Magnetic field 
		                   	Meridional Flows}

\author[addressref=Sternberg Astronomical
Institute Moscow M.V. Lomonosov State University,
email={bilenko@sai.msu.ru}]{\inits{I. 
A}\fnm{Irina A.}~\lnm{Bilenko}\orcid{0000-0002-9543-0542}}

\runningauthor{Irina A. Bilenko}
\runningtitle{Solar Polar Field Reversals}
	
\begin{abstract}
		
Based on data obtained from Wilcox Solar Observatiry the  solar polar 
magnetic fields reversals in cycles 21\,--\,25 were considered. 
The results indicate that  the polarity reversal occurs at the maximum of 
sunspot activity of each cycle, but the beginning, end, and duration of 
the reversals did not demonstrate any association 
with the Wolf numbers, which are characteristics of local magnetic fields.
Moreover, during the periods of polarity reversal, the correlation between
global magnetic field (GMF) parameters and Wolf numbers decreased and 
even moved into anti-correlation mode.

The polarity reversal occurred during periods 
of sharp structural changes in the GMF, accompanied by 
a redistribution of the positive- and negative-polarity magnetic field
domination in the North and South hemispheres.
All parameters of the GMF demonstrate characteristic 
changes associated with polarity reversal.

The polar field reversals are determined by the GMF flows of positive- 
and negative-polarity magnetic fields, which cyclically migrate from 
one pole to the opposite pole.
The magnetic fields of the new polarity are delivered  to the poles 
by a certain  flow, and then carried away by the same flow
to the opposite pole.
In each cycle, the increase in the polar magnetic field strength to its maximal values 
at the solar activity minimum and following decrees to the cycle maximum
coincides with the latitudinal changes in corresponding magnetic field flow.
The differences in start, duration, and end times of the polarity reversal 
at each pole are a consequence of the different width and speed 
of the corresponding flow.

Interaction with magnetic fields of active regions during the passage 
of GMF pole-to-pole meridional flows through low latitudes leads to the formation 
of longitudinal magnetic structures and a sectorial structure 
of the GMF.

Formulas for calculating the meridional circulation of positive- and 
negative-polarity magnetic field flows were proposed.
They allow predict the time of polarity reversals, 
and since polarity reversals occur at the maxima of cycles, then also 
the time of maxima of both the future and past cycles.
\end{abstract}
\end{frontmatter}

\section{Introduction}
\label{S-Introduction}

The solar polar magnetic fields play an important role in the
general system of solar global magnetic fields.
They have a great influence in all layers of the solar atmosphere 
and interplanetary space, their activity, structure, and cycle  variations. 
The parameters of polar magnetic fields and the time of polarity 
reversals are widely used to predict the amplitude of the next cycle
\citep{Schatten1978, Makarov1989, Choudhuri2007, Jiang2007,
Kitchatinov2011, Munoz2013, Miletskii2015, Cameron2016,
Wang2017, Upton2018, Kumar2021}.
These methods are considered to be the most physically based.
To create reliable models for predicting solar cycles, accurate data on polar 
magnetic fields,  their cyclic variations, the time 
and duration of polarity reversal are required. However, studies 
by different authors give different, and often opposite, results.
The change in the sign of the magnetic field at the poles of the Sun, 
the so-called polar field reversal, occurs  at the epoch of the sunspot 
activity maximum of each  cycle \citep{Petrie2015, Pishkalo2019}. 
Using direct observations of solar magnetic fields, the polar field 
reversal has been observed for 5 cycles already.
Observations of cycle variations of polar faculae and prominences make it 
possible to examine polarity reversal processes up to the 11th cycle
\citep{Makarov1983a, Makarov1983b, Makarov1986}.
The  polar field reversal  in the northern and southern hemispheres 
occurs at different times \citep{Makarov1983a, Makarov1983b, 
	Svalgaard2013, Pishkalo2019, Yang2024}. 
The difference can be from 0.5~yr. to a year in different cycles. 
The process of polar field reversal also differs in each cycle. 
The polarity reversal can be single or multiple in different 
cycles \citep{Makarov1983a, Makarov1983b, Makarov1986, 
Mordvinov2014, Petrie2015, Pishkalo2016, Pishkalo2019}. 
A triple polarity reversal  were found  in the 12th and 14th solar cycles in the 
South hemisphere and in the 16th, 19th, and 20th solar cycles in the North 
hemisphere  \citep{Makarov1986}.
\cite{Golubeva2023} showed the absence of multiple reversals in 
Cycles~21\,--\,24, significant differences in the time interval between reversals
in the North and South hemispheres and in the time interval between 
a reversal and the beginning of a cycle.
But \cite{Mordvinov2014} reported about a triple reversal
at the North pole in Cycle~21.
\cite{Janardhan2018} showed that the reversal in the North hemisphere 
was multiple in Cycle~24.
\cite{Pishkalo2016} and  \cite{Pishkalo2019} found  triple reversal in the North 
hemisphere in Cycle~24 and single reversal in the North and South poles
in the rest of the cases in Cycles~21\,--\,24.
\cite{Petrie2024} found that the North and South hemispheric fluxes changed 
sign multiple times during Cycle~24 polar field reversal. 
Even for one cycle, different researchers find different  
duration of the polar magnetic field reversal \citep{Sun2015,  Pishkalo2016, 
Gopalswamy2016, Janardhan2018, Petrie2024}.
Such discrepancies in the results obtained by different authors indicate 
the need for a thorough study of both the polarity reversal phenomenon itself 
and the determination of the sources of the polarity reversal process.

The study of polarity reversal is undoubtedly important 
for understanding the nature of the cyclic of solar activity and 
the creation of an adequate model of the solar dynamo.
The 'classic'  Babcock-Leighton's  model for the solar cycle
and all its modifications \citep{Babcock1961,  Leighton1964,  Leighton1969, 
Webb1984, Wang1989, Petrie2014, Sun2015, Cameron2025} 
state that the polar magnetic fields and the entire process of polarity
reversal are completely determined 
by the local magnetic fields of the  active regions (ARs).
However \cite{Fox1998} studying cycle changes in polar coronal holes (CHs) 
and polar magnetic fields in Cycles~21 and 22, concluded that the polar 
magnetic-field reversals originated from the global processes 
rather than from local magnetic field dynamics.

CHs  believed to be the good traces of the solar global magnetic field (GMF) 
dynamics and one of its characteristics such as polarity reversal
\citep{Stix1977, Webb1984, Fox1998, Bilenko2017, Harvey2002}.
A study of the polar magnetic field reversal process in Cycle~23 
using daily data on CHs in He~I~10830~\AA~line associated with 
photospheric magnetic fields  from the Kitt Peak 
Observatory, demonstrated for the first time that there is an 
anti phase migration of CHs and associated magnetic fields 
of positive and negative polarity from one pole to the opposite
that determine the solar polar field reversal  \citep{Bilenko2002}. 
Further studies using different data and different time intervals confirmed this 
result \citep{Bilenko2016, Bagashvili2017, Huang2017, Maghradze2022}.
The pole-to-pole CH migration coincides with the pole-to-pole 
meridional circulation of medium-strength photospheric magnetic fields 
\citep{Bilenko2024}. It is these magnetic fields that determine
 the process of the solar polar field reversals \citep{Bilenko2024}.
This behavior of medium-strength magnetic fields  is not consistent 
with the Babcock-Leighton theory.

It should be noted that CHs themselves are neither the source nor the cause
of the polarity reversal. They simply trace the change in the latitudinal 
distribution of the GMF. CHs follow the process of polarity reversal \citep{Bilenko2002}
with a delay  of approximately one year
in each hemisphere \citep{Webb2024}. 
Moreover, at the moment of the polarity reversals, polar CHs are absent
\citep{Stepanian2015}. 
They emphasized that  the time intervals without
CHs and the time of polar CH emergence  vary from cycle to
cycle and are different at the North and South poles of the same cycle
\citep{Stepanian2015}.
In cycles 19\,--\,23, there were intervals from 5 to 20 CRs
in which polar CHs were absent \citep{Stepanian2015}.
This means that CHs themselves are not the sources of polarity 
reversals. They merely follow changes in the GMF.

Thus, although the  solar polar magnetic fields have been studied for more than
half a century, the process  of solar polar field reversal 
and its causes are still unclear.
The purpose of this study is to investigate in detail the time-latitude and 
latitude-longitude variations in distributions and meridional flows of solar 
magnetic fields, which result in the solar polar field reversal in cycles 21\,--\,25 
as well as modeling of polarity reversal prediction. 

Section~\ref{S-data} describes the data used.
In Section~\ref{S-glob_mag_evol}, the cycle variations in the distribution
of the GMF and its structural organization in cycles 21\,--\,25 are discussed.
The changes in time and latitude-longitude distributions of positive- and
negative-polarity magnetic fields  leading to a polar field reversal
are analyzed in Section~\ref{S-pol_rev}.
The behavior of the GMF parameters during periods of polarity 
reversal is studied  in Section~\ref{S-glob_mag_field}.
The main conclusions are listed in Section~\ref{S-conclusion}.

\section{Data}
\label{S-data} 

Solar large-scale  photospheric magnetic field measurements from
WSO (Wilcox Solar Observatory) were used \citep{Scherrer1977, Duvall1977, 
Hoeksema1983, Hoeksema1984}.
WSO makes available the longest homogeneous  observational data
of the low-resolution large-scale photospheric magnetic fields
using the FeI~525.02~nm line since 1976 without major updates of their 
magnetograph \citep{Hoeksema1984, Hoeksema1986}.
At the WSO synoptic maps the radial component of the photospheric
magnetic field is shown.
No correction for the projection effect is made  because at the low 
height where the FeI~525.02~nm line is formed, the magnetic 
fields are almost radial  \citep{Hoeksema1984}.
Data on magnetic fields are given in a longitude versus sine-latitude maps
created  on the base of daily observations.
All synoptic maps consist of $30$ data points in equal steps
from $70^{\circ}$S to $70^{\circ}$N of sine latitude and
$5^{\circ}$ intervals in $0^{\circ}$ to $360^{\circ}$ longitude.
Each  map spans a full Carrington Rotation (CR, 1~CR = 27.2753 days).
The entire data set consists of 651 synoptic maps and covers CRs
1642\,--\,2292 (May 1976\,--\,December 2024).
'F-data' files with interpolation of missing observational data were used.

WSO synoptic maps of coronal magnetic field calculated from large-scale
photospheric magnetic fields with a potential field radial model with the
source-surface location at 2.5~R$_\odot$  \citep{Schatten1969,
Altschuler1969, Altschuler1975, Hoeksema1983, Hoeksema1984,
Hoeksema1986, Hoeksema1988} were used
to analyze the solar GMF dynamics.
All synoptic maps were used without any additional changes or interpolations.

The data on polar magnetic field were obtained from the WSO measurements 
of the polar photospheric magnetic fields at latitudes from about 55$^{\circ}$ 
to the pole in the northern and southern hemispheres, respectively.
Due to the projection effect, the solar rotation axis is tilted at an angle
of 7.25$^\circ$ to the Earth's ecliptic plane, the line-of-sight component 
of the photospheric magnetic fields is almost perpendicular to the radial 
field at the polar regions.
Also due to the fact that WSO daily magnetograms are taken 
with 3~arc~min resolution the polar magnetic fields in both hemispheres 
are not resolved \citep{Hoeksema1986}.
To reduce the effect of changes in the geometric projection
on the pole zone magnetic field measurements during a year,
WSO presents the polar data that are averaged over 10 days
filtered with the 20-nHz low-frequency filter.

The data on heliographic latitude of the central point of the solar disk (B0)
were taken from Kitt Peak for 1976\,--\,1979 and BASS2000 Solar Survey Archiv
of the Paris Observatory for 1980\,--\,2024.

Sunspot data from SILSO were used for comparison of GMF dynamics
with cycle variations of local magnetic fields of active regions 
\citep{SILSO_Sunspot_Number}.

\section{Global Magnetic Field Cycle Evolution in Cycles 21\,--\,25}
\label{S-glob_mag_evol} 

The reversal of the polar magnetic fields can be traced by cyclic 
variations in the hemispheric redistribution of large-scale magnetic 
fields of positive and negative polarity during solar cycles. 
The change in the sign of the polar magnetic field is most clearly manifested in the 
dynamics of magnetic fields calculated on the solar source surface.

\begin{figure}     
	\centerline{\includegraphics[width=1\textwidth,clip=]
		{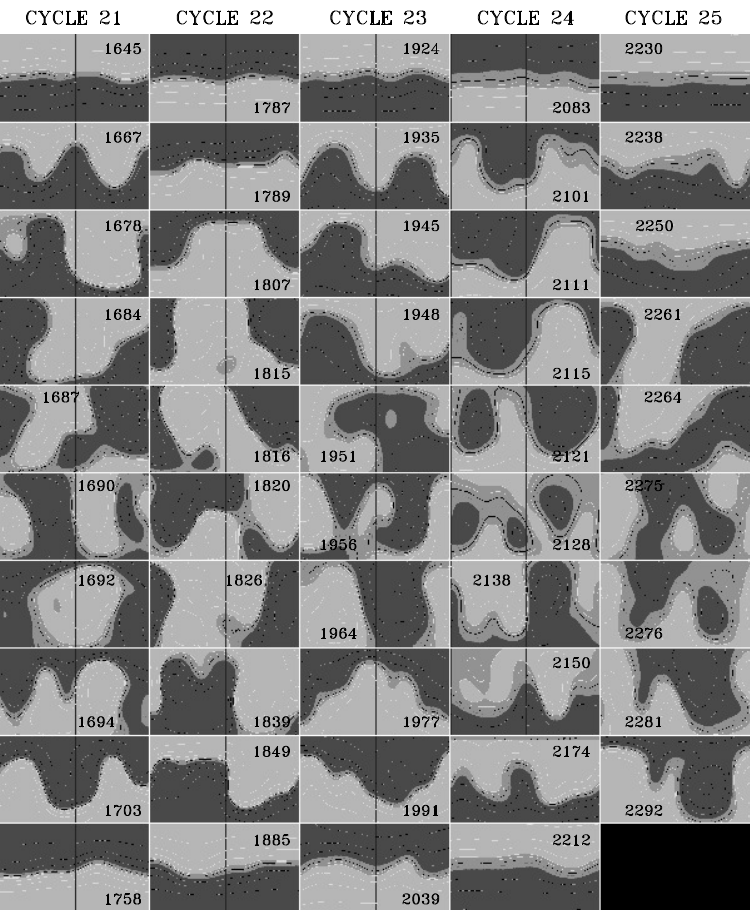}}
	\small
	\caption{Illustration of the process of transition of the GMF structure 
		from zonal at the cycle minimum to sectorial at the maximum
		and back to zonal at the minimum of the next cycle 
		with the redistribution of positive- and negative-polarity 
		magnetic field domination in the northern and southern hemispheres 
		in cycles 21\,--\,25. WSO calculated source surface synoptic maps 
		(R=2.5~R$\odot$) are shown.
		Each map is marked with a CR number.}
	\label{r25_pol_rev_iz}
\end{figure}

Figure~\ref{r25_pol_rev_iz} shows cycle variations in the structural changes 
of the solar GMF. Each column consists of WSO source surface
magnetic field synoptic maps of positive (light gray) and 
negative (dark gray) polarity for cycles 21\,--\,25. 

In Cycle~21, a clear two-sector structure was observed from CR~1978 
and it existed until CR~1685.
Starting with CR~1686, abrupt structural changes began to occur, 
as a result of which there was a redistribution of magnetic fields 
of positive and negative polarity between the 
northern and southern hemispheres and a change in the sign of the 
magnetic field at the poles to the opposite.
In CRs~1692\,--\,1694, a two-sector structure was formed for a short time, 
which then transformed into a four-sector structure that gradually 
simplified into a zonal GMF structure.
From CR~1758, GMF structure became purely zonal again.

In Cycle~22, the first signs of a sectorial structure formation
appeared in CR~1789. A clear two-sector structure was formed by 
CR~1805 and existed until CR~1815. 
Beginning with CR~1816, abrupt changes in the structure began, 
which led to a change in the dominance of the magnetic 
field polarity in each hemisphere to the opposite. 
These structural changes continued until 
CR~1821, when a two-sector structure was formed, which with 
various variations existed until CR~1936.
In CR~1937, a four-sector structure of the GMF was formed, 
which was replaced by a new two-sector structure in CR~1838.
During these reconfiguration of the GMF
the solar polar field reversal occurred.  
In this structure, there was a gradual shift of magnetic fields 
of positive polarity to the north pole, and negative polarity to the south, 
which corresponded to the new configuration of the GMF. 
This structure gradually flattened and existed until CR~1852. 
After that it began to transform into a zonal GMF structure, 
maintaining its position in longitude. An obvious zonal structure 
was observed from CR~1885.

In Cycle~23, the first signs of a sector structure appeared
in CR~1926. From CR~1932, a four-sector structure was formed, 
which with minor variations existed until CR~1943.
From CR~1944 until CR~1948  a two-sector structure was
observed.
From CR~1949, changes in the GMF structure shape  
and the  longitude  location of positive- and negative-polarity magnetic 
fields began, which continued until CR~1959 and during which there was 
a redistribution of magnetic fields of positive and negative polarity 
between the northern and southern hemispheres accompanied 
by the polar fields reversal. 
From CR~1960 to CR~1966, a two-sector structure existed
and since CR~1967 to CR~1969 -- a four-sector one.
In CR~1970, a short-lived two-sector structure was observed during one CR, 
which was replaced by a zonal GMF structure 
in the next CR~1971, which was observed up to CR~1974.
With CR~1975, a new two-sector structure 
with a different distribution of magnetic fields in longitude was formed.
Over the next 8~CRs, the GMF structure was transformed 
into a two-sector one with an opposite distribution 
of magnetic field polarities in longitude.
By CR~1955, a shift in the two-sector structure in longitude 
occurred and already in this GMF sector structure configuration 
of positive- and negative-polarity magnetic fields 
existed, gradually flattening until the transition to a zonal GMF structure. 
The zonal structure in Cycle~23 was characterized by various 
distortions, which can possibly be explained by the anomalous 
duration of the Cycle~23 decline phase.

In Cycle~24, the GMF sector structure began to form with CR~2091 and, 
with some modifications, existed until CR~2109. Then, from CR~2110,
the longitude distribution changed somewhat, although the main 
distribution of positive- and negative-polarity magnetic fields
by longitude remained the same until CR~2015, when a four-sector
structure was formed. This four-sector structure with various complex 
reconfigurations existed until CR~2132.
A new two-sector structure was formed starting with CR~2133 and existed 
until CR~2138. During these periods, the process of redistribution 
of positive and negative polarity magnetic fields between the northern 
and southern hemispheres and the polar field reversal occurred, 
continuing until CR~2148. 
Then a variable structure was observed that turned into a four-sector 
structure in CR~2174, which then gradually simplified to a zonal one.

In Cycle~25, the GMF sector structure began to form with CR~2238. 
The two-sector structure formed in CR~2250 and lasted until CR~2261. 
Starting with CR~2264, sharp structural changes began with 
the appearance of a new two-sector structure during 
CR~2266\,--\,2272 and further reorganization of the positive- and 
negative-polarity magnetic field distribution  in the northern and southern 
hemispheres of the Sun accompanied by the polar magnetic field sigh
change in each hemisphere. Structural changes 
continued until CR~2292, i.e. the end of the data under consideration.

In each cycle, there was a transition from a zonal distribution 
of magnetic fields at the minimum of solar activity to a sectorial one 
at the maximum, when a redistribution of the dominance of large-scale 
positive- and negative-polarity magnetic fields in each hemisphere 
and the solar polar field reversal occurred. 
During declining phases, the distribution of GMF again becomes 
zonal, but with a polarity opposite to that at the beginning of the cycle. 
Thus, in all the cycles considered, the polarity reversal occurred during periods 
of sharp structural changes in the GMF, accompanied by a redistribution 
of the dominance of large-scale magnetic fields of positive and negative 
polarity between the northern and southern hemispheres.
In general, the polarity reversal process follows the same scenario 
in each cycle.The period of these cyclic changes
is equal, on average, to 11 years. Thus, the full magnetic cycle 
of the GMF is $\approx$22 years.

\section{Solar Polar  Magnetic Field Reversals in Cycles 21\,--\,25}
\label{S-pol_rev} 

\begin{figure}     
	\centerline{\includegraphics[width=1\textwidth,clip=]
		{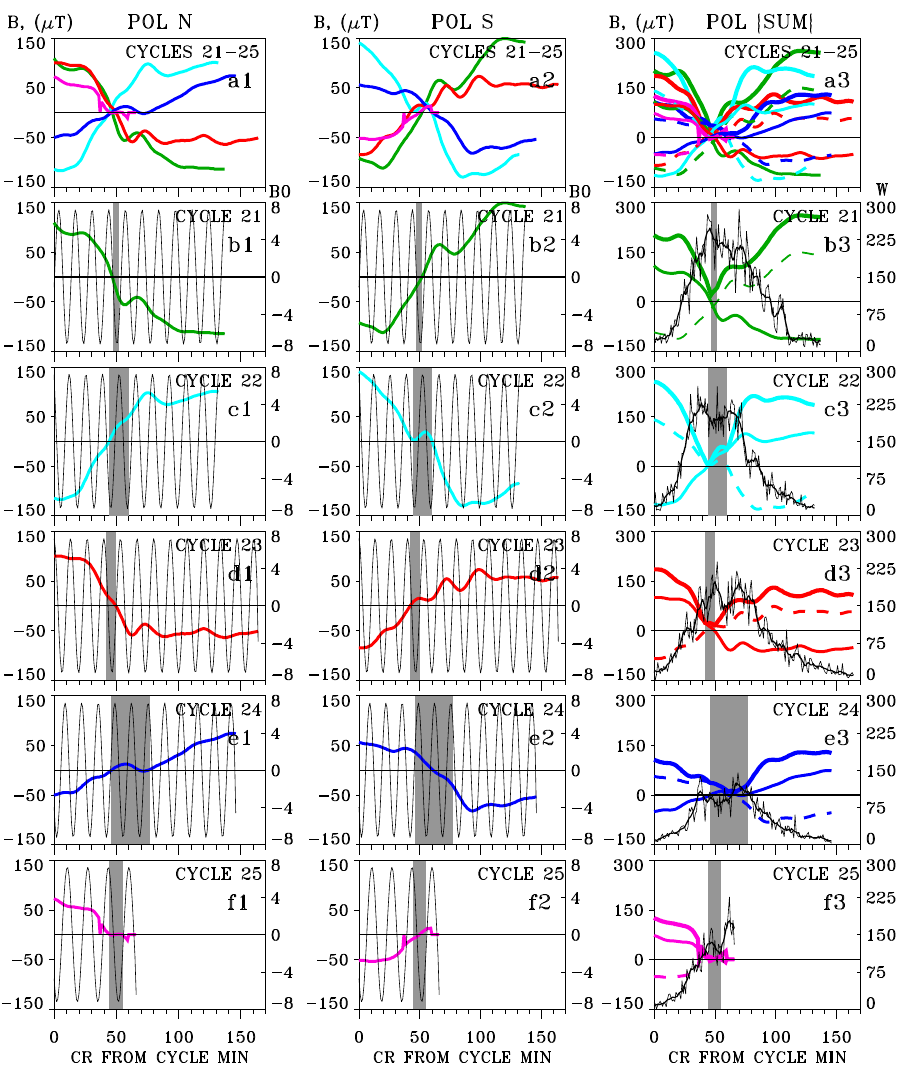}}
	\small
	\caption{Polar magnetic field strength evolution in each cycle separately 
		for the North (b1\,--\.f1) and South (b2\,--\.f2) poles and their sum (a1, a2).
		(b3\,--\,f3) combined North (solid line) and South (dashed line) polar 
		magnetic fields and the sum of their moduli (thick lines) for each cycle 
		and for all cycles (a3).}		
	\label{pol_n_s_sum}
\end{figure}

Figure~\ref{pol_n_s_sum} shows the changes in the polar magnetic 
field strength in each cycle separately for the North (first column) 
and South (second column) poles, their combination and the sum 
of their moduli (third column). 
The CR averaged Wolf numbers are shown by black lines  for each cycle in
Figures~\ref{pol_n_s_sum}(b3\,--\,f3, right 'y'-axes). 
The periods of the polar field reversals are highlighted in gray. 
They were defined based on the time of the magnetic field sign change
at the North and South poles (Figures~\ref{pol_n_s_sum}(b1\,--\,f1 and b2\,--\,f2).
It should be noted that due to the inclination of the solar rotation axis 
to the Earth's  orbit by 7.25$^\circ$, the polar magnetic field cannot 
be observed all the time. 
The North solar pole is tilted towards us during September and South pole is
tilted towards us during March of each year. The maximum areas of the polar 
regions can be observed only during these periods.
Accordingly, the accuracy of determining the moment of polarity reversal in each 
hemisphere cannot be better than half a year. Thus, we can judge the change 
in the dominance of magnetic fields of positive or negative 
polarity at the pole only by studying near-polar zones at high latitudes.
The changes in the angle of  polar zone seen from the Earth (B0) are shown
in black lines in Figures~\ref{pol_n_s_sum}(b1\,--\,f1 and b2\,--\,f2, right 'y'-axes).
Some characteristics of the polar field reversals and Wolf numbers 
in Cycles 21\,--\,25 are summarized in Table~1.
$Wmax$ is the value of the Wolf numbers at the cycle maxima, 
and $Wmin$ is that at the cycle minima. $TWmax$ is the number of the CR
corresponding to the maximum of the Wolf number, 
and $TWmin$ that to the minimum. 
$\triangle Tcycle$ is the cycle duration. 
$Tbeg$ is the CR at the beginning of the polarity reversal, 
and $Tend$ is that at its end.
$B0beg$ is the range of $B0$ during the CR at the beginning 
of the polarity reversal, and $B0end$ is that at its end. 
$\triangle T$ is the number of CRs from the cycle minimum 
to the beginning of the polarity reversal.

Comparison of the polar magnetic field reversal characteristics
with the Wolf numbers, reflecting variations in local magnetic fields, 
indicates the absence of any relationship between them.
Although in odd cycles the duration of the polarity reversal is less
than in even cycles, the polarity reversal parameters do not depend 
neither on the magnitude nor on duration of the cycle in which the 
reversal occurs (see Figure~\ref{pol_n_s_sum} and Table~\ref{Tab_pol}).
Though the polarity reversals always 
take place at the maximum of sunspot activity, 
they do not coincide unambiguously in time with a specific peak 
in Wolf numbers. In Cycle~21, the polarity reversal 
began after the first peak of spot activity and continued during 
the decline of this peak. 
In Cycle~22, the polarity reversal occurred during the period 
of the Gnevyshev's gap, i.e. between the first and second peaks 
of the Wolf numbers. 
In Cycle~23, the polarity reversal began before the first 
peak of the Wolf numbers and continued until the maximum 
of this peak. In Cycle~24, the polarity reversal was the 
longest for the entire observation period. It began at the 
decline of the Wolf number peak and covered both the 
Gnevyshev's gap and the second spot activity peak. 
In  Cycle~25, the polarity reversal period coincided with the 
first peak of the Wolf numbers.
In each cycle, the polarity reversal began $\approx$42\,--\,45~CRs after the 
date of corresponding cycle minimum specified in the SILSO database.

\begin{table}
	\caption{Solar Cycles 21\,--\,25 and polar reversal parameters.}
	\label{Tab_pol}
	\begin{tabular}{lccccc}     
		\hline                    
		Parameter & \multicolumn{5}{c}{Cycles}    \\
		&   21   & 22 & 23 & 24 & 25 \\
		\hline
		Wmax                                       & 232.9 & 212.5 &  180.3 & 116.4 &    -   \\
		Wmin                                        & 17.8   & 13.5   &  11.2   &  2.2    &  1.8    \\
		TWmax, (CR)                 &  1690 &  1822 & 1983   &  2149  &   -    \\
		TWmin, (CR)                  & 1639  & 1780  & 1913   & 2078   & 2225 \\
		$\triangle$Tcycle, (CR)   &   141  &   133  &  165   &   147   &   -    \\
		Tbeg, (CR)                    & 1689  & 1824   & 1955   & 2124   &  2269 \\
		B0beg, (deg.)     &  0.98  & -2.70:-5.48  & 6.16:3.76 & -1.41:1.70 &-6.91:-5.29 
		\\
		Tend, (CR)                    & 1694   & 1840  & 1963    & 2155  & 2280  \\
		B0end, (deg.)     & -5.56  & -7.23:-6.27 & -2.36:0.74 &7.17:5.93 & -4.83:-6.73 \\
		$\triangle$T, (CR)                     &   5     &  16     &   8       &  31    &   11   \\
		T from min, (CR)                       &  50    &   44   &   42     &    46  &   44  \\
		\hline
	\end{tabular}
\end{table}

According to Figure~\ref{pol_n_s_sum}, both at the North and South 
poles there were a one-time transition to the dominance of a new 
magnetic field polarity except the South pole in Cycle~22 and  
North pole in Cycle~24, where the transition to the new polarity 
magnetic field dominance had an oscillatory character 
(Figures~\ref{pol_n_s_sum}(c2, e1)).
The duration of the polarity reversal was shorter in the odd Cycles~21, 23, 
and 25 and significantly longer in the even Cycles~22 and 24. 
In Cycle~24, the polar field reversal was the longest
approximately  from CR~2124 to CR~2155.

\begin{figure}     
	\centerline{\includegraphics[width=1\textwidth,height=1\textheight,clip=]
		{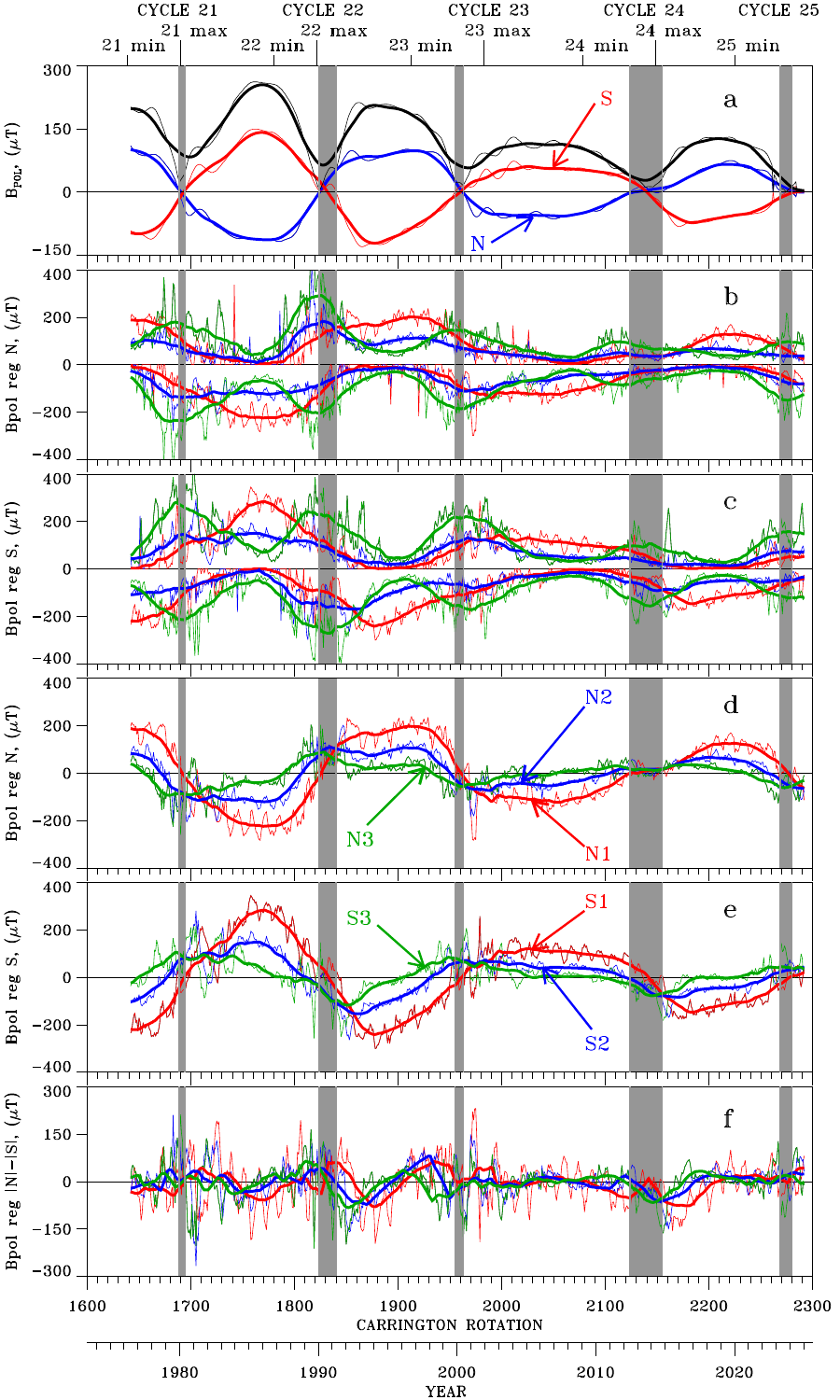}}
	\small
	\caption{(a) variations in the  polar magnetic 
		fields at the North (blue) and South (red) poles in cycles 21\,--\,25. 
		The sum of their moduli is shown in black;
		(b) variations of the positive- and negative-polarity magnetic field strength 
		in the latitudinal ranges:
		45$^\circ$\,--\,55$^\circ$ are shown in green, 
		55$^\circ$\,--\,65$^\circ$ -- in blue, 
		and 65$^\circ$\,--\,70$^\circ$ -- in red in the North hemisphere
		and (c) that in the South hemisphere; 
		(d)  mean magnetic field strengths for each latitudinal interval  (N1\,--\,N3)   
		in the North hemisphere and 
		(e)  that in the South hemisphere   (S1\,--\,S3);     
		(f) imbalances between the mean magnetic field strength 
		in the North and  South hemispheres for each latitudinal interval. 
		Thin lines indicate the magnetic fields averaged over each CR, 
		and thick lines that over seven CRs.  
		The maxima and minima of Cycles 21\,--\,25 are marked at the top.}
	\label{pol_field_reg}
\end{figure}

Figure~\ref{pol_field_reg}(a)  shows variations in the CR-averaged 
(thin lines) and seven CR-averaged (thick lines) polar magnetic 
field strength at the North (blue) and South (red) poles in Cycles~21\,--\,25. 
The sum of their moduli is shown in black. 
The asymmetry of polar magnetic fields between the northern and southern 
hemispheres was more pronounced in high Cycles~21 and 22.

Due to angle $B0$ variations, changes in magnetic fields the 
near polar zones can only be examined in detail, when studying 
variations in polar magnetic fields over a long time interval.
Using the H-alpha synoptic charts for the period 1904\,--\,1982 and 
the data on the polar prominences for the period 1870\,--\,1905,
Makarov \citep{Makarov1983a, Makarov1983b, Makarov1986} revealed 
poleward migration of the magnetic neutral line in the latitudes 
from 50$^\circ$ to 90$^\circ$ in both the hemispheres.
The poleward drift velocity of the neutral line varied from 
4.2 to 8.2~m~s$^{-1}$ in Cycles~14\,--\,17  \citep{Makarov1983b}
and from 5.4 to 13.4~m~s$^{-1}$ in Cycles~18\,--\,21 with a peak 
of 29.4~m~s$^{-1}$ in the northern hemisphere in Cycle~20 
\citep{Makarov1983a}.
They also noted that at low polar 
latitudes (50$^\circ$), polar reversal occurs earlier than at high
latitudes (90$^\circ$).
For Cycles~21\,--\,24 \cite{Pishkalo2019} found that polarity reversal 
in the near-polar latitude range $\pm$(55$^\circ$\,--\,90$^\circ$) 
occurred 0.5\,--\,2.0 years earlier than the time when the reversals were 
completed in the corresponding pole.
\cite{Yang2024} using the Hinode/SP high spatial resolution 
polar observations from 2012 to 2021 revealed that
the magnetic field polarity in each solar polar reversed from the
70$^\circ$ latitude to the pole successively from low to high latitudes
at the epoch of solar maximum.
They concluded that this indicates the presence of the magnetic flux migration 
from low latitudes to the pole.

Figures~\ref{pol_field_reg}(b\,--\,f) show variations in 
large-scale positive- and negative-polarity photospheric magnetic 
fields in the near-polar zones in different latitude ranges. 
From Figures~\ref{pol_field_reg}(b, c), it follows that in the periods before 
the polar field reversal, the magnetic field strength gradually increased 
from low to high latitudes, in each latitudinal interval at the North 
(Figure~\ref{pol_field_reg}(b)) and South (Figure~\ref{pol_field_reg}(c)) poles.
At lower latitudes in the range of 45$^\circ$\,--\,55$^\circ$ (green), 
where the influence of ARs is still felt, this is less clearly evident.
Figures~\ref{pol_field_reg}(d\,--\,e) show more clearly that magnetic fields 
of the new polarity first begin to dominate at lower latitudes in the range 
of 45$^\circ$\,--\,55$^\circ$ (green), 
then at latitudes of 55$^\circ$\,--\,65$^\circ$ (blue), 
and only then at the highest latitudes of 65$^\circ$\,--\,70$^\circ$ (red) 
which is consistent with the results obtained by 
\cite{Makarov1983a}, \cite{Makarov1983b}, \cite{Makarov1986}, 
\cite{Pishkalo2019},  and \cite{Yang2024}.
Thus, the polarity reversal occurs from low to high latitudes.
At the same time, there is an imbalance between the magnetic field 
strength in the northern and southern hemispheres for the same
latitude ranges (Figure~\ref{pol_field_reg}(d)).

\begin{figure}     
	\centerline{\includegraphics[width=1\textwidth,clip=]
		{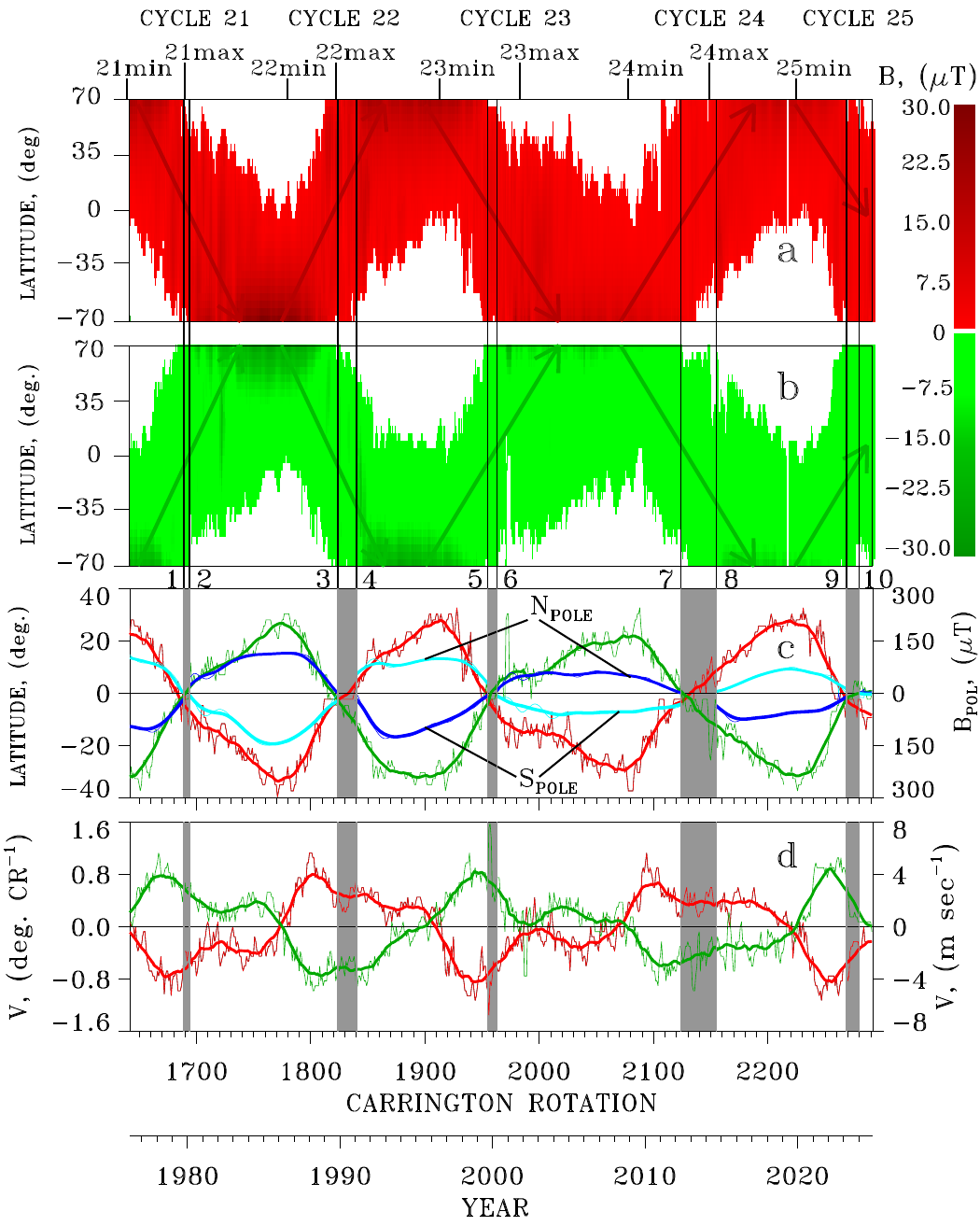}}
	\small
	\caption{Cyclic variations of large-scale positive- (red, (a)) and negative-polarity 
		(green, (b)) magnetic fields calculated on the source surface; 
		(c) the trajectories of the flow centers (red and green respectively).
		Variations in the positive polarity (dark blue) and negative polarity (light blue) 
		polar magnetic field strength at the North (N) and South (S) poles.
		On the right 'y'-axes the strength of the polar magnetic field is given
		in absolute values;
		(d) speed of displacement by latitude of positive- and negative-polarity 
		magnetic field flows.
		The maxima and minima of Cycles 21\,--\,25 are marked at the top.}
	\label{mag_pol_rev}
\end{figure}

But it's also interesting to see what happens to magnetic fields 
at different latitudes afterwards. According to the Babcock-Leighton theory, 
magnetic fields should somehow migrate beneath the photosphere 
at high latitudes.
However, Figures~\ref{pol_field_reg}(b\,--\,e) show that after a polarity 
reversal at each pole, the magnetic field strength reached the maximum values 
in each latitudinal interval, which, with minor variations, persisted until 
the next polarity reversal at that pole. 
At the next polarity reversal, the magnetic field strength 
began to decrease and  change the magnetic field polarity to the opposite
in each latitudinal range. 
These changes also went from low polar latitudes to high ones.
This indicates that the magnetic fields of the new polarity were delivered 
to the poles by a certain meridional flow, and then 
carried away by the same flow.

For a closer look at the flows in Figure~\ref{mag_pol_rev}, 
the time-latitude distributions of longitude-averaged positive-polarity 
magnetic fields (Figure~\ref{mag_pol_rev}(a)) and in that of the negative 
polarity  (Figure~\ref{mag_pol_rev}(b)) calculated on the source surface 
in the range from 0.6 to 30 $\mu$T are presented.
Red and green arrows in Figures~\ref{mag_pol_rev}(a and b) show 
the directions of  the meridional flows.

These large-scale meridional flows occupied a wide range 
of latitudes  in each CR that can reach $\approx$70$^\circ$\,--\,100$^\circ$.
The flows were antiphase and antisymmetric with respect 
to the equator. They are a manifestation of the meridional circulation
of the large-scale positive- and negative-polarity magnetic fields
in each cycle \citep{Bilenko2024}.
It should be emphasized that these magnetic-field flows show the dynamics 
of magnetic fields calculated on the source surface that reflects the cycle 
variations of the solar GMF. The magnetic fields of ARs 
are absent from source surface synoptic maps.
It should be noted that these magnetic field flows have a 
complex structure. They do not represent a continuous movement 
of solar plasma on the surface of the Sun, but appearing successively 
shifting in latitude the zones of dominance of positive- or 
negative-polarity magnetic fields.
The trajectories of the centers of these flows were different in 
different cycles (red and green lines in Figure~\ref{mag_pol_rev}(c)).
As the values of flow mean latitudes are very chaotic and changed 
strongly from one CR to the next (Figure~\ref{mag_pol_rev}(c, red and 
green thin lines)), they were smoothed by 41~CRs (thick lines).
The mean latitudes of the magnetic field flows of positive and
negative polarity changed in anti phase, their variations in different cycles
were different, their maximum values were significantly shifted 
in time relative to each other  (Figure~\ref{mag_pol_rev}(c)). 
According to Figures~\ref{mag_pol_rev}(a, b) a new polarity magnetic field 
was carried by the corresponding flow toward the opposite poles. 
Due to the large width of the stream at the onset of the polarity reversal, 
a significant portion of it is still located at low latitudes, near the equator
and in the opposite hemisphere. Therefore, the latitude-averaged value 
is located near the equator.

Thus, the magnetic fields of the new polarity do 
not remain in the polar zone, but were carried away by these flows.
Such behavior of positive- and negative-polarity 
magnetic fields indicates the existence of a constant, 
periodic, oppositely directed large-scale flows of positive- and
negative-polarity magnetic fields from one pole to the opposite with 
a period of about 22~years.
Such magnetic flows correspond to the dynamics of large-scale 
medium-strength photospheric magnetic fields \citep{Bilenko2024}.

For comparison in Figure~\ref{mag_pol_rev}(c), variations in the polar 
magnetic field strength at the North (N) and South (S) poles are also shown, 
taking into account the polarity of the field.
From Figure~\ref{mag_pol_rev}(c), it follows that the increase in 
the magnetic field strength of the new polarity at each pole is uniquely 
correlated with an increase in the latitude localization of the magnetic flux 
of the same polarity. The growth of the polar magnetic 
field intensity in each polar zone coincides with the growth in 
latitude of the corresponding magnetic field flow.
Thus the increase in magnetic field strength at the poles was determined 
by newly arriving portions of the magnetic field.
The growth was provided by the influx of new magnetic field brought by the 
flows of each polarity to the corresponding pole.

The polarity reversal at each pole began after the new polarity magnetic flux  
reached $\approx$70$^\circ$ latitude at the corresponding pole.
These moments are marked by black vertical lines numbered 1, 2, 3, ...,10
in Figure~\ref{mag_pol_rev}.
The different start, duration, and end times of the polarity reversal 
in the northern and southern hemispheres are a consequence 
of the different width and speed of the positive- and negative-polarity 
magnetic field flows.
After the bulk of the flux reached maximal latitudes, 
a reverse magnetic flow migration began toward the opposite pole. 
At the same time, the magnetic field intensity in the polar zone began 
to decrease (Figure~\ref{mag_pol_rev}(c)).
Thus, cyclic variations of the polar magnetic field, including polarity 
reversal, are completely determined by the pole-to-pole
meridional flows of the GMF.

The magnetic field was carried away from one pole to the opposite one, 
and this determined a decrease in magnetic field strength at that pole.
This indicates that these magnetic flows 
that migrate from pole to pole determine not only the time of the polarity reversal, 
but they are the source of the polar magnetic field strength growth to the cycle 
minimum  and then its decrease to the maximum of the next cycle. 
These large-scale magnetic field flows coincide well with the cyclic pole-to-pole
migration found for CHs \citep{Bilenko2002, Bilenko2016, Huang2017, 
Maghradze2022} which serves as a confirmation that it is these magnetic fields
that determine the process of the solar polar field reversals.

\cite{Bagashvili2017} studying the rotation of coronal holes, came to the 
conclusion that the CH rotation latitudinal characteristics do not match 
any known photospheric rotation profile. 
According to their results, the CH rotation profile coincides the lower layers 
of convection zone at around 0.71~R$_{\odot}$ and with tachocline.  
They concluded that it is possible that CHs are associated with the GMF,
which originates in the tachocline region.
This may be evidence that the polarity reversal is determined by processes 
occurring in the tachocline, at the base of the convective zone.

Figure~\ref{mag_pol_rev}(d) shows the velocity changes of the flows. 
Variations in the velocity of positive- and negative-polarity magnetic fields   
were not symmetrical.
The velocities were higher at low latitudes and decreased to zero 
at sunspot minima when the meridional flows turn around in polar regions.
Despite the velocity of the flows was rather low, about 1$^\circ$ per 
CR or 7~m~sec$^{-1}$, the large width of flows allowed 
them to transport a new flux to the corresponding pole during a cycle
for the polar field reversals.
The obtained velocity values coincide with that were determined in 
\cite{Makarov1983a, Makarov1983b, Makarov1986} for 
the neutral line movements in the near polar regions.
It should be noted that  the velocity of the large-scale magnetic-field flows 
did not depend on neither the height nor the duration of a solar cycle 
determined by the Wolf numbers.

\begin{figure}     
	\centerline{\includegraphics[width=1\textwidth,clip=]
		{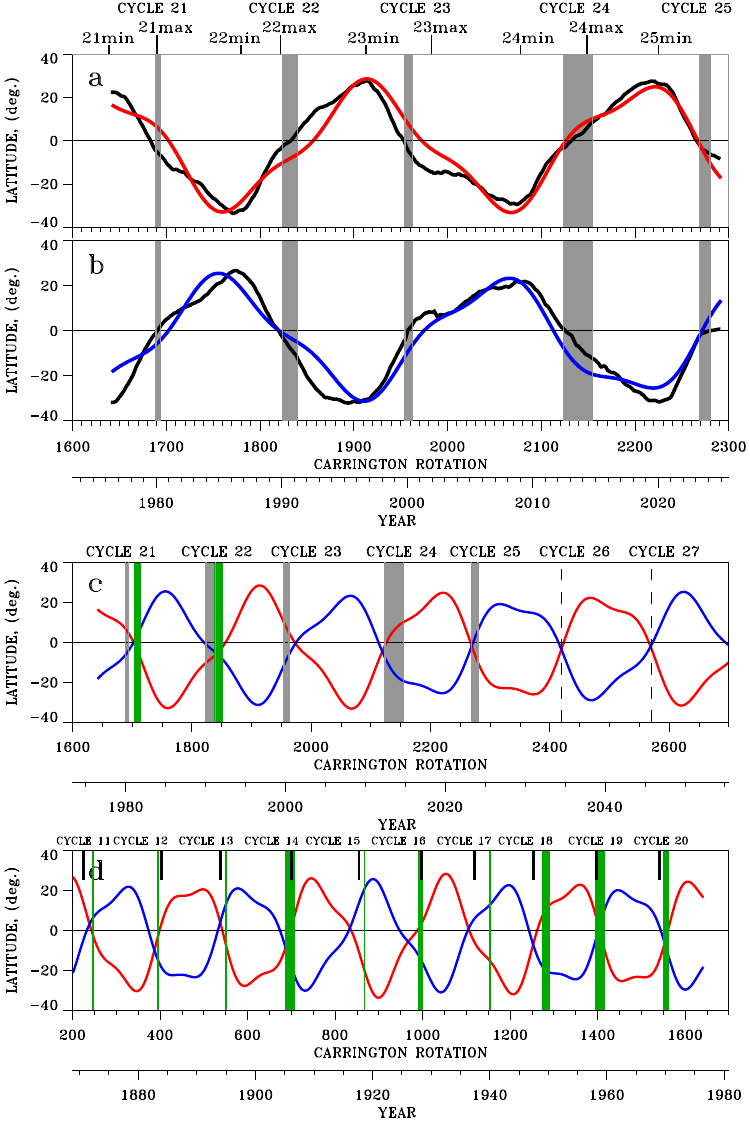}}
	\small
	\caption{Time-latitude distribution of  the flow centers of longitude-averaged        
		         positive-polarity (a) and negative-polarity (b) magnetic field flows;
		     (c) forecast of the polarity reversal time in Cycles~26 and 27;
		     (d) estimated time of polarity reversals in Cycles~11\,--\,20.}
	\label{lat_pol_progn}
\end{figure}

The cyclical changes in mean latitudes of the positive- and
negative-polarity magnetic field flows look like sinusoid and cosinusouid,
with some distortions at low latitudes (Figure~\ref{mag_pol_rev}(c)). 
These distortions may be due to the influence of magnetic fields of
ARs. Although ARs themselves lack magnetic fields 
on the source surface by definition, their interaction with magnetic flows 
at the photosphere and subphotosphere levels, when the flows passe through 
the latitudes of ARs, can influence the speed and shape 
of magnetic flows, leading to the observed distortion of the sinusoids.
The shape of the curve of each flow is fairly well approximated 
by two sinusoids using the formula for positive-polarity magnetic field flow:
\begin{eqnarray} 
	L{^\circ}_{pos}(CR) = -2.8 - 26 \times cos(t/45.9 - 0.87) + 5.3 \times 
	cos(t/17.05 + 0.87)
	\label{eq_pos}	
\end{eqnarray}
and for negative-polarity magnetic field flow:
\begin{eqnarray}
	L{^\circ}_{neg}(CR) = -3 - 24 \times sin(t/46.2 - 1.97) + 4.7 \times sin(t/17.25 
	+ 0.5)
	\label{eq_neg}	
\end{eqnarray}
\noindent where $L$ is latitude in degrees and $t$ is time in CR number.

In Figure~\ref{lat_pol_progn}(a), the time-latitude distribution of the 
flow centers of positive-polarity magnetic fields (red) and in 
Figure~\ref{lat_pol_progn}(b) that of the negative polarity (blue) 
calculated on the source surface and approximated (black lines) using 
the formula (\ref{eq_pos}) for positive- and (\ref{eq_neg}) for negative-polarity 
magnetic field flows are presented.
The periods of the flows are slightly different
for positive- and negative-polarity magnetic fields.
This leads to different start and end times of polarity reversals 
at the North and South poles in the same cycle.

The formulas allow us to predict the time
of polarity reversals in the following cycles.
Figure~\ref{lat_pol_progn}(c) shows the variations in the location centers 
of positive- and negative-polarity magnetic field flows predicted using  
formulas \ref{eq_pos} and \ref{eq_neg} in Cycles 26 and 27, 
indicated by vertical dashed lines. 
It is still difficult to say how accurate this forecast is, but
these formulas can also be used to determine the time 
of polarity reversals in past cycles.
Figure~\ref{lat_pol_progn}(d) shows modeled variations 
in large-scale magnetic field flows of positive (red) and negative (blue) 
polarity in Cycles 11\,--\,20. 
Since polarity reversals occur at the cycle maxima, this makes 
it possible to check the correctness of such a forecast of the  polarity
reversal time. In Figure~\ref{lat_pol_progn}(d), the cycle 
maxima are indicated  by thick black lines at the top. 
The data on cycle maxima were taken from the WDC-SILSO database.
Taking into account that the formulas for the forecast were determined only 
using two 22-year observed magnetic cycles, the coincidence can be
considered good.
Since polarity reversals always occur at the peaks of solar activity, 
forecasting using these formulas allows us to determine the approximate 
time of the maxima of subsequent cycles.
The data on the time of polarity reversal obtained by 
\cite{Makarov1986, Makarov1994} from the study of prominences and 
faculae can also serve as a check. 
The time of polarity reversal obtained by them in cycles 11\,--\,22  are
marked in green in Figures~\ref{lat_pol_progn}(c and d).
The coincidence of the polarity reversal times calculated using
formulas \ref{eq_pos}~and~\ref{eq_neg} and that from 
\cite{Makarov1986, Makarov1994} is quite good, and in a number 
of cycles it is a complete coincidence.

\begin{figure}     
	\centerline{\includegraphics[width=1\textwidth,clip=]
		{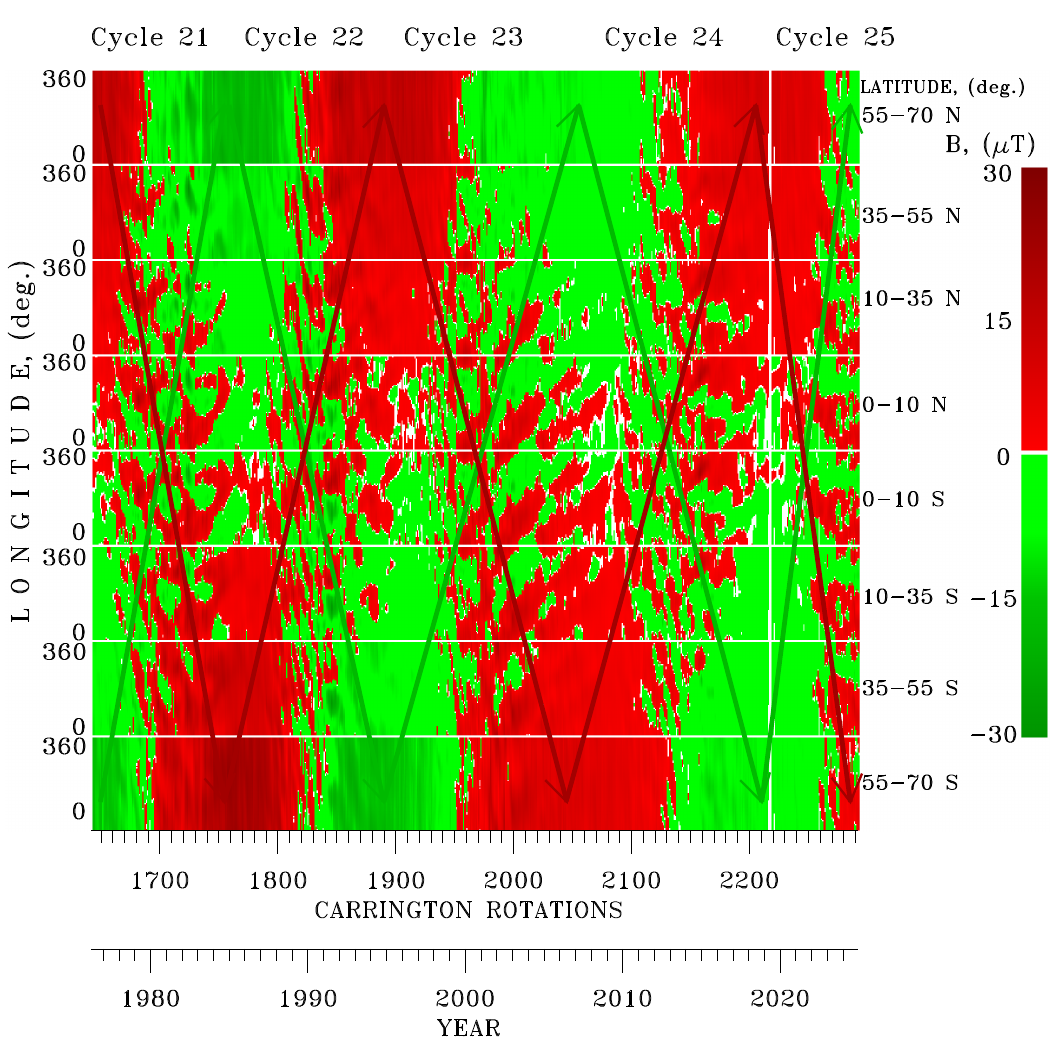}}
	\small
	\caption{The time-longitude distributions of magnetic fields calculated 
		on the source surface for different latitude intervals.}
	\label{r25_lon_lat}
\end{figure}

\begin{figure}     
	\centerline{\includegraphics[width=1\textwidth,clip=]
		{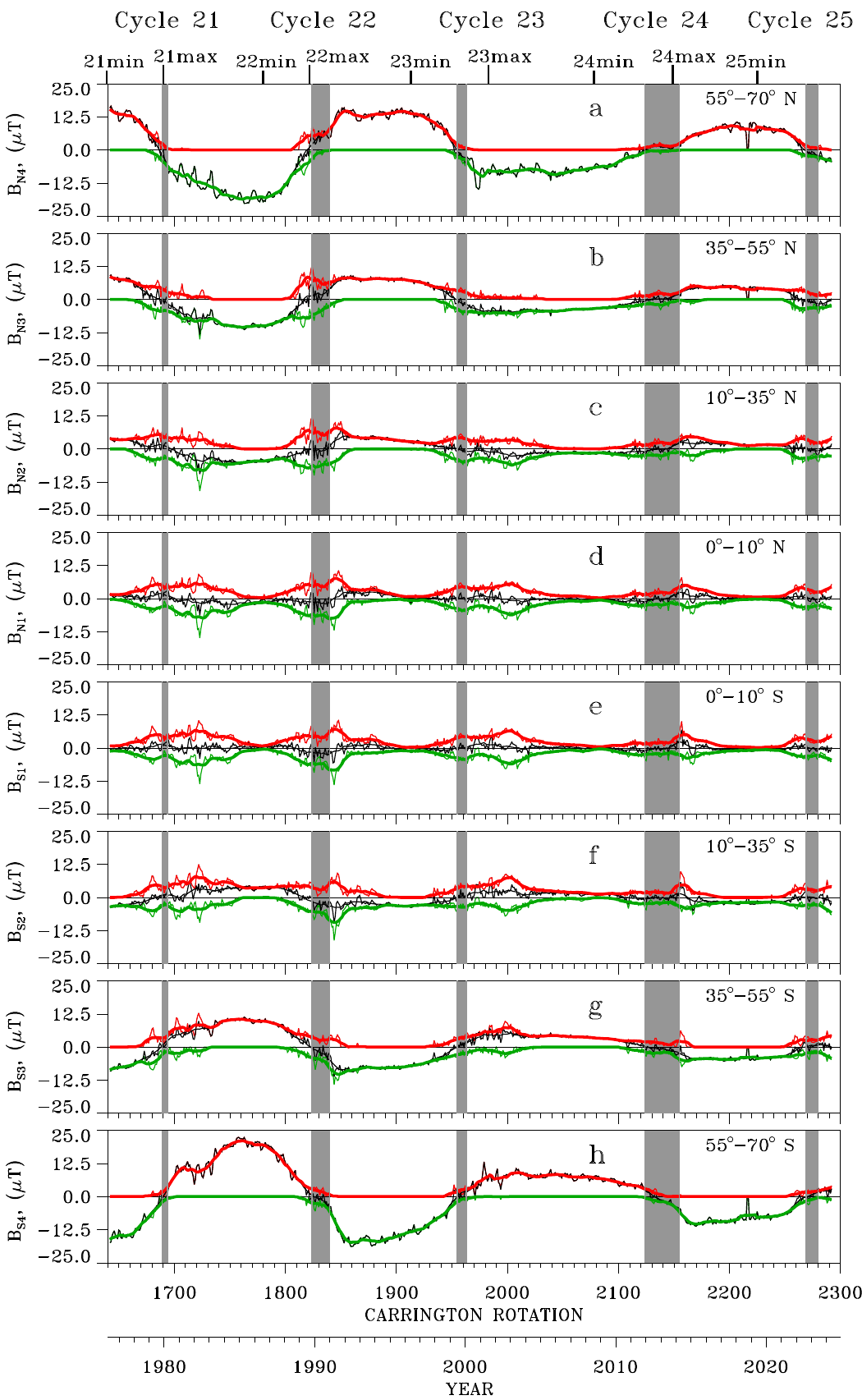}}
	\small
	\caption{Variations in magnetic field strength in different latitude intervals from
		their	longitude-time diagrams calculated 	on the source surface.}
	\label{r25_mag_n_s}
\end{figure}

Using the chromospheric Ca~II~K polar network data within 
the latitude range of $\pm$(55$^{\circ}$\,--\,90$^{\circ}$) for both 
the northern and southern hemispheres, \cite{Mishra2025} 
reconstructed the historical polar magnetic field from 1904 to 2022.
A comparison of the polarity reversal times in the data they 
obtained for positive- and negative-polarity magnetic field variations and those 
calculated using formulas~\ref{eq_pos}~and~\ref{eq_neg} shows 
very good agreement (see for example Figure~3 in \cite{Mishra2025}).
The good agreement between the polarity reversal times obtained 
using formulas~\ref{eq_pos}~and~\ref{eq_neg} and sunspot maxima 
dates from SILSO database and polar reversal times
revealed  by \cite{Makarov1983a, Makarov1983b, Makarov1986} 
and \cite{Mishra2025} indicates the 
possibility of using these formulas to predict the time of polar magnetic 
field reversals and the periods of maxima of subsequent cycles.

When constructing time-latitude distributions of magnetic fields 
(Figure~\ref{mag_pol_rev}(a, b)), 
as with the classic butterfly diagram, information about the distribution
of magnetic fields by longitude is completely lost. However, variations 
in the time-longitude distributions of magnetic fields contain a wealth 
of information, including information related to polarity reversal 
(Section~\ref{S-glob_mag_evol}, Figure~\ref{r25_pol_rev_iz}).
It was shown that in the latitude range from -55$^\circ$ S to +55$^\circ$ N, 
large-scale magnetic fields are structured by longitude  
\citep{Bilenko2002, Bilenko2014, Bilenko2016}.
The time-longitude distributions of magnetic fields calculated on the source 
surface for latitude intervals of 0$^\circ$\,--\,10$^\circ$, 10$^\circ$\,--\,35$^\circ$, 
35$^\circ$\,--\,55$^\circ$, and 55$^\circ$\,--\,70$^\circ$ in the North and South 
hemispheres are shown in Figure~\ref{r25_lon_lat}. 
Changes in the magnetic field distribution that reflect the changes in the 
sign of the polar magnetic fields are visible. Magnetic field polarity changes 
are clearly traced in individual longitudinal intervals. 
If we consider the evolution of magnetic fields of each polarity in time, 
at all latitudes then sinusoidal flows of positive- and negative-polarity 
magnetic fields are revealed. They are indicated by arrows in a color 
corresponding to the polarity of the magnetic field.

The structured longitudinal distribution of magnetic fields may be 
a consequence of the interaction of the GMF flows 
as they migrate from pole to pole with the low-latitude magnetic fields 
of ARs.
As GMF flows pass through low latitudes, i.e., AR latitudes, 
their magnetic fields interact with the magnetic fields of the ARs. 
This leads to a change in the GMF flows velocity and a distortion 
of their shape, as demonstrated above  (Figure~\ref{mag_pol_rev}), 
as well as to the formation of longitudinal structures. 
This corresponds to the formation of the GMF sectorial structure 
in each cycle  (Figure~\ref{zon_sect_zs}).
From Figure~\ref{r25_lon_lat} it follows that the sectorial structure 
was formed at latitudes of the ARs' locations.

In Figure~\ref{r25_mag_n_s}, the CR mean values 
of positive- (red lines) and negative-polarity (green lines) magnetic fields  
and their imbalance (black lines) calculated from 
Figure~\ref{r25_lon_lat} for each latitudinal interval are presented.
In each cycle, the increase and decrease of the imbalance of positive- 
and negative-polarity magnetic fields reflecting the cyclic process 
of polarity reversal is clearly traced from one pole to the opposite pole.

\section{Characteristics of the Global Magnetic Field during
	              Solar Polar Field Reversals}
\label{S-glob_mag_field} 

\begin{figure}     
	\centerline{\includegraphics[width=1\textwidth,clip=]
		{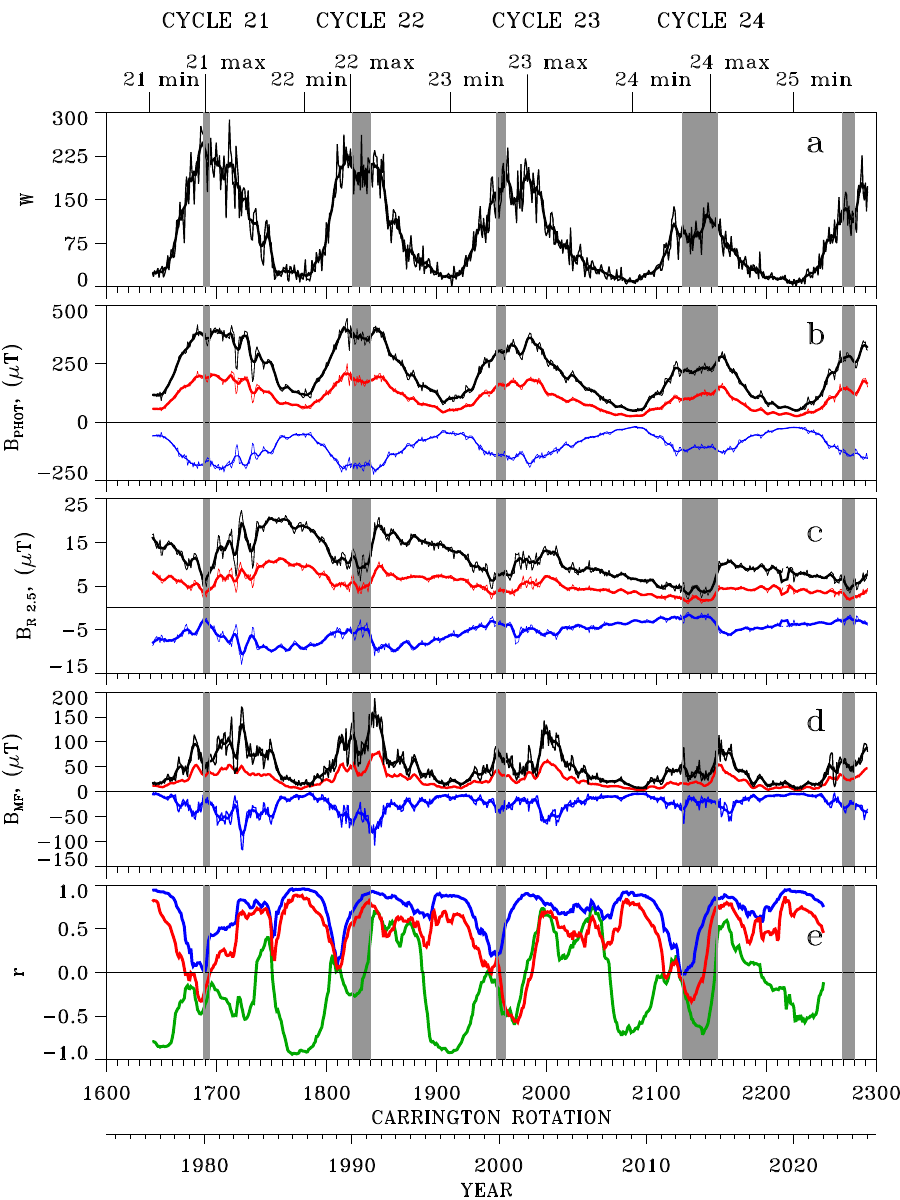}}
	\small
	\caption{Cycle variations in CR-averaged (thin lines) and 
		          7 CR-averaged (thick lines):
		(a) Wolf  numbers;
		(b) large-scale photospheric magnetic fields;
		(c) source-surface magnetic fields calculated at 2.5~R$\odot$;
		(d) magnetic field of the Sun as a star;
		(e) correlation coefficients between W and B$_{PHOT}$ (blue),
		W and B$_{R~2.5}$ (green),  W and B$_{MF}$ (red).
		In (b\,--\,d) positive-polarity magnetic fields are indicated in red
		and negative-polarity magnetic fields are indicated in blue.
		Polarity reversal periods are indicated in gray.
		The maxima and minima of Cycles 21\,--\,25 are marked at the top.}
	\label{din_cor_art}
\end{figure}

Figure~\ref{din_cor_art} shows a comparison of cycle variations of Wolf numbers 
(W) characterizing the dynamics of local magnetic fields and such 
characteristics of the GMF as the photospheric large-scale magnetic field 
of the Sun (B$_{\rm PHOT}$), the magnetic field calculated on the source 
surface (B$_{\rm R~2.5}$), and the magnetic field of 
the Sun as a star  (B$_{\rm MF}$).
In Figure~\ref{din_cor_art}(e), the variations in correlation between 
B$_{\rm PHOT}$ and W (blue), B$_{\rm R~2.5}$ and W (green),  
B$_{\rm MF}$  and W (red) for each 41~CRs (3.06~yr) with 1~CR step 
are presented.
During the periods of polarity reversal (marked in gray) and preceding it, 
the correlation decreased.
For correlation between  B$_{\rm MF}$ and W (red) in cycles 21, 23, and 24
and for correlation between B$_{\rm R~2.5}$ and W (green),
it even moving on anti-correlation mode. 
The maximal negative correlation between B$_{\rm R~2.5}$ and W (green)
was during minima and rising phases in each cycle
at the time were no ARs, and the polar magnetic field, which is one of the 
characteristics of the GMF, had maximum values in each cycle.

Spherical harmonic analysis is often used to investigate the solar GMF.
In spherical analysis, the magnetic field is described as a function of 
latitude and longitude coordinates ($r, \theta, \phi$)  by the potential function
\citep{Chapman1940, Altschuler1969, Altschuler1975, Altschuler1977}:
\begin{eqnarray}
 \psi(r,\theta,\phi)=R \sum_{n=1}^N \sum_{m=0}^n
\left(\frac{R}{r}\right)^{n+1} [g_n^m \, {\rm cos}\left(m\phi\right)\, + \, h_n^m \, {\rm 
sin}\left(m\phi\right)] \, P_n^m(\theta), 
     \label{eq_psi}
\end{eqnarray}
\noindent where  $P_{n}^m(\theta)$ are the associated Legendre
polynomials, and $N$ is the number of harmonics. The coefficients
$g_n^m$, $h_n^m$ are calculated using a least mean-square fit to
the observed line-of-sight photospheric magnetic fields with a
potential field assumption. 
The potential-field model is useful for analyzing the basic properties 
of the GMF, such as its symmetric and asymmetric components, 
dipole, quadrupole, multipole characteristics, tilt, etc., which allow us 
to better understand the solar activity behavior over different cycles.
The different harmonic power spectra can be
calculated \citep{Altschuler1977, Levine1977}:
\begin{eqnarray}
	S_n=\sum_{m=0}^n [(g_n^m)^2 + (h_n^m)^2].
	\label{eq_spectr}
\end{eqnarray}
The temporal evolution of the lowest order (n = 1, 2, 3, . . ., 9) harmonics
at the outer-boundary radius R = 2.5~R$_{\odot}$ were considered
The higher-order harmonic power spectra follow the sunspot cycle
and  are dominated by local magnetic fields of ARs.

\begin{figure}     
	\centerline{\includegraphics[width=1\textwidth,clip=]
		{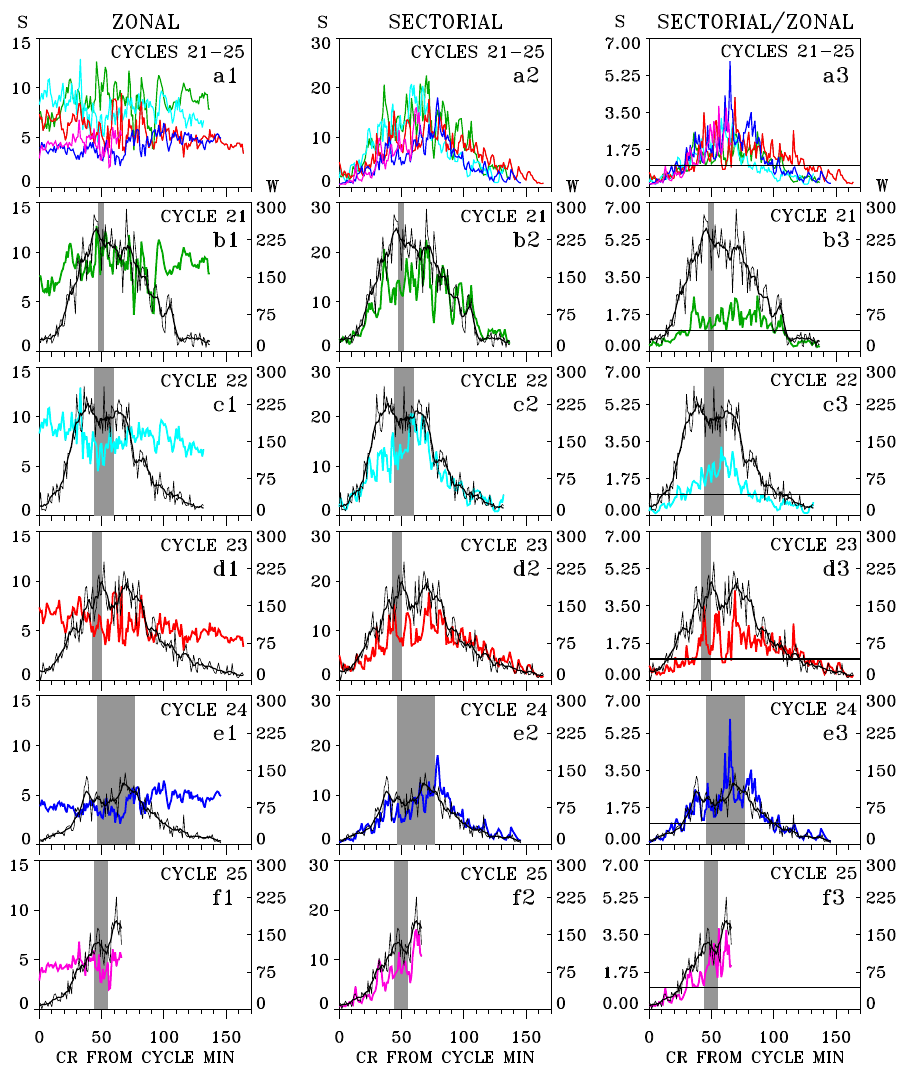}}
	\small
	\caption{(b1\,--\,f1) evolution of zonal harmonics sum in each cycle and
		their combined graph (a1).
		(b2\,--\,f2) evolution of sectorial harmonics sum in each cycle and
		their combined graph (a2).
		(b3\,--\,f3) evolution of the ratio of the sectorial to zonal harmonic sum 
		in each cycle and their combined graph (a3).
		The horizontal lines in (a3\,--\,f3) marks the level where the sum 
		of the sectorial harmonics is equal to that of the zonal ones.
		Polarity reversal periods are indicated in gray.}
	\label{zon_sect_zs}
\end{figure}

\begin{figure}     
	\centerline{\includegraphics[width=1\textwidth,clip=]
		{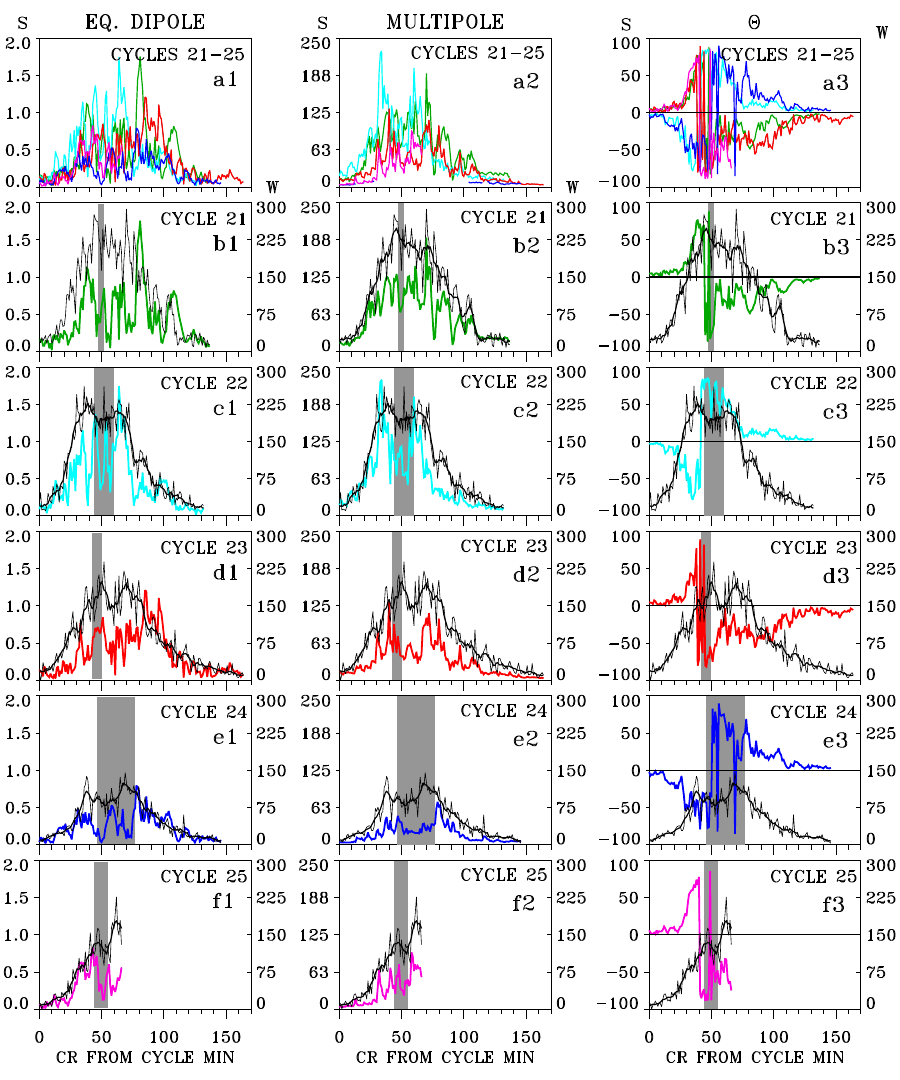}}
	\small
	\caption{(b1\,--\,f1) evolution of zonal harmonics sum in each cycle and
		their combined graph (a1).
		(b2\,--\,f2) evolution of multipoles sum in each cycle and
		their combined graph (a2).
		(b3\,--\,f3) evolution of the ratio of the sectorial to zonal harmonic sum 
		in each cycle and their combined graph (a3).
		Polarity reversal periods are indicated in grey.}
	\label{edip_mult_teta}
\end{figure}

\begin{figure}     
	\centerline{\includegraphics[width=1\textwidth,clip=]
		{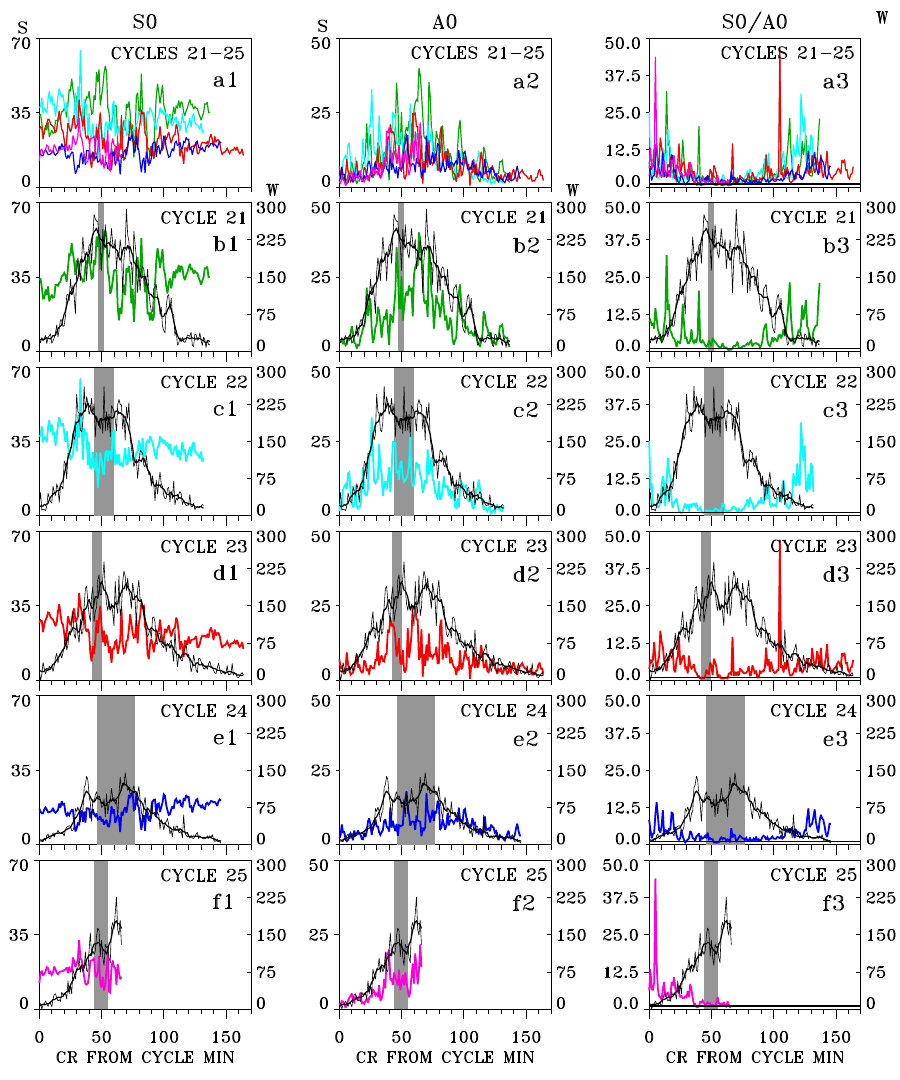}}
	\small
	\caption{(b1\,--\,f1) evolution of the axisymmetric, 
		and symmetric  with respect to the equator harmonics (S0)  in each cycle 
		and their combined graph (a1);
		(b2\,--\,f2) evolution of axisymmetric, but antisymmetric  with respect 
		to the equator (A0) harmonic in each cycle and their combined graph (a2);	
		(b3\,--\,f3) evolution of the polar angle, $\theta$, of the dipole component
		in each cycle and their combined graph (a3).
		Polarity reversal periods are indicated in grey.}
	\label{sim_nonsim_simnonsim}
\end{figure}

The zonal harmonic spectra sum is presented in Figures~\ref{zon_sect_zs}(b1\,--\,f1) 
and that of sectorial harmonics is shown in Figures~\ref{zon_sect_zs}(b2\,--\,f2).
The ratio of the sectorial to zonal harmonic sum is displayed in 
Figures~\ref{zon_sect_zs}(b3\,--\,f3). 
Their combined spectra are shown in Figures~\ref{zon_sect_zs}(a1, a2, a3)
respectively.
For comparison with the sunspot activity, each graph shows variations 
in the Wolf CR-averaged  (thin black lines) and  7~CR-averaged
(thick black lines) W numbers.

From Cycle~21 to Cycle~25, the maximal values of both zonal
and sectorial harmonics decreased.
Sector harmonics had maximum amplitudes at the maxima of each cycle, 
but near the polarity reversal period there was a local decrease in their values
in each cycle.
The amplitudes of the sum of zonal harmonics were distributed more evenly 
throughout the cycles with some variations, but near the polarity reversal periods
their values slightly decreased. 
During periods of polarity reversal, the ratio of the sectorial to zonal 
harmonics decreased, indicating a decrease in the role of 
sectorial and an increase in zonal harmonics during polar reversals.
In Cycles~21 and 22, the excess of the sum of sectorial harmonics 
over the sum of zonal harmonics was insignificant. 
Beginning with Cycle~23, the excess increased significantly and 
reached its maximum value in Cycle~24.
This indicates an increasing role of sectorial harmonics in weak cycles
(Figures~\ref{zon_sect_zs}(b3\,--\,f3)).

Similar dependencies are presented for the variations 
in equatorial dipole (\ref{eq_eqv_dip}),
the sum of multipoles (\ref{eq_spectr}), and $\theta$ angle (\ref{eq_theta}) 
for individual Cycles~21\,--\,25 (Figures~\ref{edip_mult_teta}(b1\,--\,f1, 
b2\,--\,f2, and b3\,--\,f3)) and that combined for all cycles 
(Figures~\ref{edip_mult_teta}(a1\,--\,a3))
\begin{eqnarray}
	S_{eqv.dip.} =   \sqrt{(g_1^1)^2 + (h_1^1)^2}
	\label{eq_eqv_dip}
\end{eqnarray}
\begin{eqnarray}
	\tan\theta=(g_1^0)^{-1} [(g_1^1)^2 + (h_1^1)^2]^{1/2}.
	\label{eq_theta}
\end{eqnarray}
The equatorial dipole and the sum of multipole harmonics had
maximum values  at the maximum of each cycle.
However, during periods of polarity reversal, their values decreased.
From Figures~\ref{edip_mult_teta}(a3\,--\,f3), it follows that the sign 
change in each cycle began after significant jumps in the polar angle. 
During the polarity reversal period, the deviations of the polar angle
reached maximum values and were oscillatory.
In each cycle after polarity reversal, sharp deviations of the polar angle 
ceased and its values gradually decreased to zero.

In Figure~\ref{sim_nonsim_simnonsim} the axisymmetric and symmetric 
with respect to the equator harmonic spectra sum (S0, (\ref{eq_s0}), 
Figures~\ref{sim_nonsim_simnonsim}(a1\,--\,f1))
\begin{eqnarray}
	S0=\sum_{n=2,4,6,8}(n+1)g_n^0  P_n(\theta)
	\label{eq_s0}
\end{eqnarray}
and the axisymmetric and antisymmetric  with respect to the equator
harmonic spectra (A0, (\ref{eq_a0}), Figures~\ref{sim_nonsim_simnonsim}(a2\,--\,f2))
\begin{eqnarray}
	A0=\sum_{n=1,3,5,7,9}(n+1)g_n^0 P_n(\theta)
	\label{eq_a0}
\end{eqnarray}
defined according to \cite{Stix1977} are presented.
The  evolution of their ratio are shown in 
Figures~\ref{sim_nonsim_simnonsim}(a3\,--\,f3).
At the maximum of each cycle, S0 decreased.
A0 component had maximum values at the maximum of each cycle, 
but during polarity reversal its values decreased.
During periods of maximum activity and polarity reversals, the ratio of 
S0 to A0 components decreased sharply in each cycle.
This indicates a diminished role of S0 components of the GMF 
during these periods and increasing role of  the  axisymmetric 
and antisymmetric with respect to the equator components of the GMF.

\begin{figure}     
	\centerline{\includegraphics[width=1\textwidth,clip=]
		{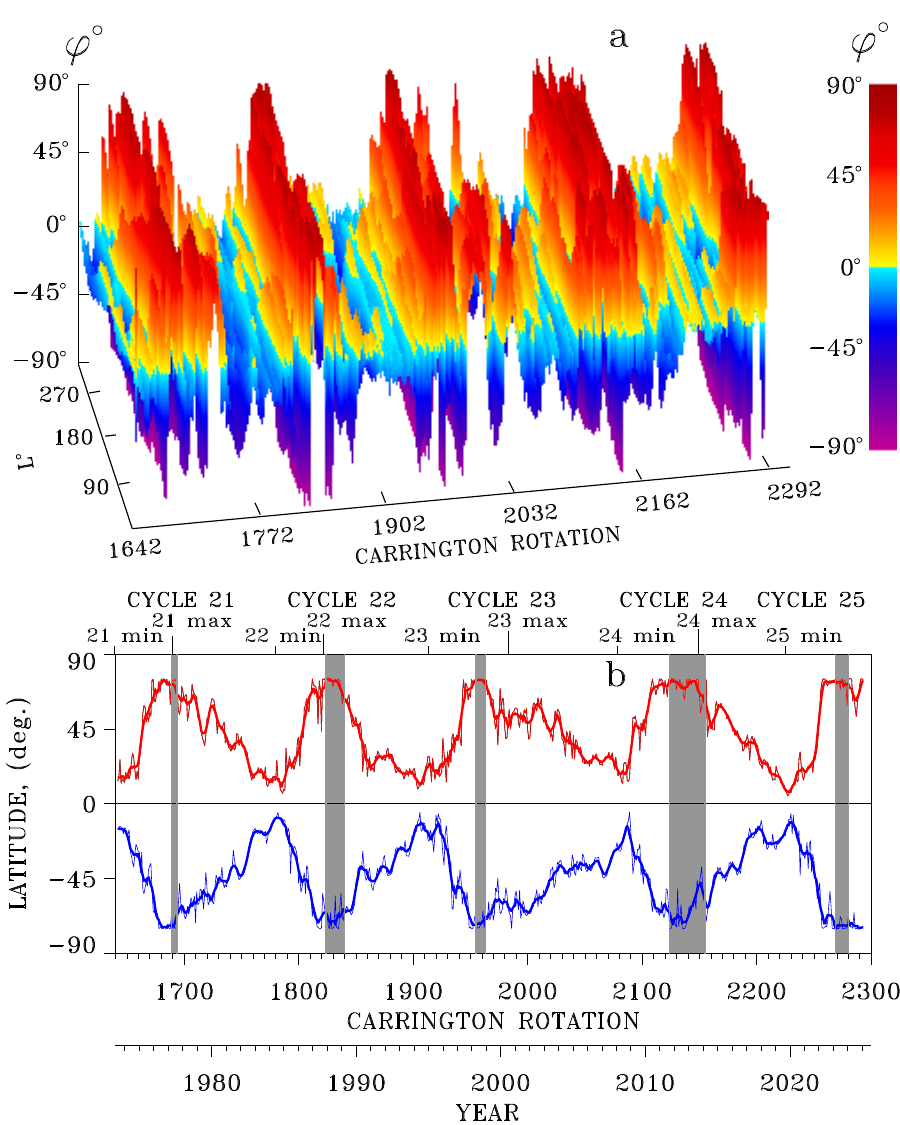}}
	\small
	\caption{(a) Cycle variations in the location of the neutral line 
		(base of the heliospheric current sheet);
		(b)  the latitude distribution of the heliospheric current sheet 
		in the northern and southern hemispheres in cycles 21\,--\,25.
		Solar polar field reversal periods are denoted in gray.}
	\label{neit_line_hcs}
\end{figure}

One of the characteristics of the GMF is also the position 
of the neutral field line  (Figure~\ref{neit_line_hcs}(a)), i.e. the base
of the heliospheric current sheet (HCS,  Figure~\ref{neit_line_hcs}(b)).
At the minimum of solar activity, with a zonal 
structure domination of the GMF, the HCS is located near the solar equator. 
With the growth of activity, its configuration changes, 
reflecting the formation of a sectorial structure (Figure~\ref{zon_sect_zs}). 
The length of the neutral line underlying the 
HCS increases in latitude and at the maximum of activity can reach 
70$^\circ$ in latitude in each hemisphere.
The growth in HCS latitude location coincides with the increase 
in $\theta$ angle variations (Figure~\ref{edip_mult_teta}) 
and domination of antisymmetric and antysimmetric with respect to 
the equator (A0) components  (Figure~\ref{sim_nonsim_simnonsim}).

\cite{Pishkalo2019} found that reversals occurred one\,--\,two years after 
maximal HCS tilts  in Cycles 21\,--\,24.
In Cycles 21\,--\,25 the structural changes associated with the solar polar field 
reversals were also revealed in cyclic variations of the neutral
line  (Figure~\ref{neit_line_hcs}(a)), 
the base of the HCS (Figure~\ref{neit_line_hcs}(b)). 
It follows from Figure~\ref{neit_line_hcs} that the time of the 
polar magnetic field reversals coincided with the periods of maximum
deviations in the latitude of the HCS.
In each cycle, the polarity reversal began after the neutral line 
reached its maximum latitudes.

It should be noted that the maximum latitude position of the HCS 
did not depend neither on the height nor on duration of the cycle 
determined by Wolf numbers and remained at the same high level 
in all the cycles under consideration.

\section{Conclusion} 
\label{S-conclusion} 

Although the polarity reversal occurred at the maximum of sunspot 
activity in Cycles 21\,--\,25, the beginning, end, and duration of polarity reversal
did not demonstrate any association with the Wolf numbers, 
which are characteristics of local magnetic fields.
They did not depend on either the duration or the height of 
the cycle determined by the Wolf numbers.
Moreover during the periods of polarity reversal, the correlation between
GMF parameters (B$_{\rm PHOT}$, B$_{\rm R~2.5}$,  and B$_{\rm MF}$)
and Wolf numbers decreased and even moved into anti-correlation mode.
The maximum latitude position of the GMF neutral line, the base of HCS, 
did not also depend neither on the height nor on duration of an AR  cycle 
and remained at the same maximal level during polar reversals in all cycles.

The polarity reversal occurred during periods 
of sharp structural changes in the GMF, accompanied by 
a redistribution of the dominance of large-scale magnetic fields of positive 
and negative polarity between the northern and southern hemispheres.
In cycles 21\,--\,25, the polarity reversal began $\approx$42\,--\,45~CRs 
after the cycle minimum.

The reversal of the magnetic field is determined by GMF flows of positive- 
and negative-polarity magnetic fields, which cyclically migrate from 
one pole to the opposite pole.
The increase of the polar magnetic field strength in each polar zone 
coincides with the growth in latitude of the corresponding magnetic field flow
and is determined by the magnetic field brought by this flow.
The polarity reversal at each pole began after the magnetic flux of the 
new polarity reached 70$^\circ$ latitude at the corresponding pole.
After the bulk of the flux reached maximal latitudes, 
a reverse magnetic flow migration began toward the opposite pole. 
At the same time, the magnetic field intensity in the polar 
zone began to decrease.
Thus, the magnetic fields of the new polarity are delivered  to the poles 
by a certain meridional flow, and then carried away by the same flow
to the opposite pole.

The differences in start, duration, and end times of the polarity reversals 
in the northern and southern hemispheres of the same cycle are a 
consequence of the different width and speed of the positive- 
and negative-polarity magnetic field flows.

Formulas for calculating the latitudinal circulation of positive- and 
negative-polarity magnetic flows were proposed.
These formulas allow us to predict the time of polarity reversals, 
and since polarity reversals occur at the maxima of cycles, then also 
the time of maxima of both the future and past cycles.

During the polarity reversals the magnitude of 
zonal and sectorial harmonics, $A0$ and $S0$ components, equatorial
dipole, and multipoles locally diminished, angle $\theta$  and HCS 
had maximal oscillation amplitudes. 
The excess of sectorial harmonics over the zonal ones during the 
maxima phases increased in low cycles and reached 
the maximum in Cycle~24.
This indicates an increasing role of sectorial harmonics in weak cycles.
But during periods of polarity reversal, the ratio of the sectorial 
to zonal harmonics decreased. This indicates a decrease in the role of 
sectorial and an increase in zonal harmonics during polar reversals.
During the periods of polarity reversals, the role of axisymmetric and 
symmetric relative to the equator GMF components decreased and
the role of  axisymmetric and antisymmetric with respect to the equator 
components of the GMF increased.

As magnetic flows pass through low latitudes, 
moving from pole to pole, their magnetic fields interact with 
the magnetic fields of ARs. 
This leads to a distortion of their shape and velocity, as well as to 
the formation of longitudinal structures. 
This corresponds to the formation of a sectorial GMF structure in each cycle.

The regularities found in the variations of the parameters 
of the solar GMF during periods of polar magnetic field polarity reversals and 
the association of meridional GMF flows that determine the change 
in the sign of the magnetic field at the solar poles with the dynamics 
of coronal holes may indicate that the polarity
reversal is determined by processes occurring in the tachocline, at the 
base of the convective zone but this remains to be investigated.

%
\begin{acks}
The study was conducted under the state assignment of 
Lomonosov Moscow State University.

Wilcox Solar Observatory data used in this study was obtained via the web site
$http://wso.stanford.edu$ at $2025:08:21_06:53:24$ PDT courtesy of J.T. Hoeksema.

The data on heliographic latitude of the central point of the solar disk (B0)
were taken from Kitt Peak for 1976\,--\,1979 and BASS2000 Solar  Survey Archiv
of the Paris Observatory for 1980\,--\,2024.

Sunspot data from the World Data Center SILSO, Royal Observatory of 
Belgium were used.
Source: WDC-SILSO, Royal Observatory of Belgium, Brussels, \\
DOI: $https://doi.org/10.24414/qnza-ac80$.
\end{acks}



\bibliographystyle{spr-mp-sola}
\bibliography{bilenkoarchv}  

\end{document}